\documentstyle[aps,prl,epsfig]{revtex}
\begin{document}
\draft
\twocolumn[\hsize\textwidth\columnwidth\hsize\csname %
@twocolumnfalse\endcsname

\title{On the electron-energy loss spectra and plasmon resonance in
cuprates}
\author{P. Prelov\v sek$^{1,2}$ and P. Horsch$^3$  }
\address{ $^{1}$J. Stefan Institute, 1000 Ljubljana, Slovenia }
\address{ $^{2}$Faculty of Mathematics and Physics, University of
Ljubljana, 1000 Ljubljana, Slovenia }
\address{$^{3}$Max-Planck Institut f\"ur Festk\"orperforschung,
Heisenbergstr. 1, D-70569 Stuttgart, Germany}
\date{\today}
\maketitle
\begin{abstract}\widetext
The consequences of the non-Drude charge response in the normal state
of cuprates and the effect of the layered structure on electron-energy
loss spectra are investigated, both for experiments in the
transmission and the reflection mode. It is shown that in the
intermediate doping regime the plasmon resonance has to be nearly
critically damped as a result of the anomalous frequency
dependence of the relaxation rate. This also implies an unusual
low-energy dependence of the loss function. Both facts are consistent
with experiments in cuprates. Our study based on the $t$-$J$ model
shows good agreement with measured plasmon frequencies.

\end{abstract}
 ]
\narrowtext

In metals the investigation of the plasmon resonance is used as a
standard tool to extract the information about charge carriers. As
several other electronic properties also the plasmon response is quite
unusual in cuprates. This can be realized from the dielectric
measurements \cite{tann} which reveal a very broad plasmon resonance
in the electron-energy loss (EEL) function and hence a poorly defined
plasma frequency $\omega_p$. One of the experimental approaches to
measure the plasmon dispersion $\omega_p({\bf q})$ is the
electron-energy loss spectroscopy (EELS) \cite{pine}. So far the EELS
has not been employed very frequently in cuprates, also due to the
ambiguities in the interpretation. The EELS in the transmission mode
has been performed mostly for Bi$_2$Sr$_2$CaCu$_2$O$_8$ (BSCCO)
\cite{nuck,wang}, showing an optic plasmon with a quadratic
dispersion. Similar spectra, restricted to $q=0$, have been obtained
for La$_{2-x}$Sr$_x$CuO$_4$ (LSCO) by analysing the dielectric
reflectivity \cite{uchi}. Recently higher-resolution EELS spectra
have been measured \cite{schu} for BSCCO in the reflection mode,
displaying an acoustic-like plasmon at small wavevectors ${\bf q}$. In
both experiments the plasmon resonance appears heavily damped,
although not overdamped.

So far there have been only few theoretical studies of EEL, i.e. relevant
to the quite specific electronic structure of cuprates. It is has been
realized early, in analogy with the layered electron gas \cite{fett}, that
weakly coupled layers of mobile electrons in CuO planar structures can
lead in addition to optic plasmons also to acoustic plasmon branches
\cite{grif}. Recently, an important relation of the EEL function and
the related dielectric function $\epsilon({\bf q},\omega)$ to the
mechanism of the superconductivity in cuprates has also been stressed
\cite{legg}.

Another important aspect follows from the observation that the charge
dynamics in the normal state of cuprates is anomalous and not consistent with
the Drude-type relaxation. Charge as well as spin fluctuations have
been instead phenomenologically described within the
marginal-Fermi-liquid (MFL) scenario \cite{varm,litt}. In particular,
the latter has been shown to apply in the optimum-doping regime to the
optical conductivity $\sigma(\omega)$, which can be expressed within a
generalized Drude form with an effective (MFL-type) relaxation rate
$\tau^{-1} \sim 2\pi \lambda (\omega+\xi T)$ \cite{litt}. More
recently, the same anomalous dynamical behavior has been found in
numerical studies within the $t$-$J$ model at intermediate doping
\cite{jakl1,jakl2,jakl3}, revealing a more precise form of this
novel diffusion behavior.
Closely related to the dielectric function $\epsilon({\bf
q},\omega)$ and the EELS in cuprates are  calculations of the
density fluctuations $N({\bf q},\omega)$, studied within the $t$-$J$
model both analytically \cite{khal} and numerically \cite{tohy,jakl3}.

The aim of this contribution is to discuss the consequences of the
established non-Drude planar optical conductivity for the loss function
and EELS experiments, both in the transmission and in the reflection
mode. In particular we shall investigate the intrinsic damping of the
plasmon mode.

The microscopic models, such as the Hubbard model and the $t$-$J$ model
\cite{rice},
studied in connection with electronic
properties of cuprates and other materials with strongly correlated electrons do
not incorporate the long-range Coulomb interaction. The latter is
necessary for the appearance of plasmon oscillations. It is quite
straightforward to include the long-range interaction within the
random-phase approximation (RPA), where the
dielectric function $\epsilon({\bf q},\omega)$ and the dynamical
density susceptibility $\chi({\bf q},\omega)$ can be represented as
\begin{eqnarray}
\epsilon({\bf q},\omega)&=&1+V_{\bf q}\chi_0({\bf q},\omega), \nonumber
\\ \chi({\bf q},\omega)&=&{\chi_0({\bf q},\omega) \over 1+V_{\bf
q}\chi_0({\bf q},\omega)}, \label{eq1}
\end{eqnarray}
where $\chi_0({\bf q},\omega)$ is the electron density suceptibility
which includes the short-range correlations.  In
cuprates the hopping-matrix element for electrons between layers is
rather small, as e.g. manifested in the large resistivity anisotropy, so
that $\chi_0({\bf q},\omega)$ should depend
only on the planar component ${\bf q}_\parallel$, i.e. $\chi_0({\bf
q}_\parallel,\omega)$. On contrary, the Coulomb
matrix element $V_{\bf q}$ still depends on the 3D wave vector
${\bf q}=({\bf q}_\parallel,q_z)$.  The general expression for $V_{\bf q}$ 
has been given in Refs.\cite{fett,grif}. For
simplicity we restrict our discussion to small ${\bf q}$, i.e. $q_z
d <1, q_\parallel a < 1 $, where $d, a$ denote the interlayer
distance and the planar cell dimension, respectively. In this regime
one gets
\begin{equation}
V_{\bf q}=e^2/\epsilon_0(\epsilon_{\parallel}
q_{\parallel}^2+\epsilon_{\perp} q_z^2),\\
\label{eq2}
\end{equation}
where $\epsilon_{\parallel}$ and $\epsilon_{\perp}$ are  planar and
interlayer $\omega \to \infty$ dielectric constants, respectively,
due to screening of non-conduction electrons.
The loss function is given by \cite{pine}
\begin{equation}
I({\bf q},\omega)= -{\rm Im} {1\over \epsilon({\bf q},\omega)}=
V_{\bf q} {\rm Im} \chi({\bf q},\omega). \label{eq3}
\end{equation}
In the following
we shall focus on the long-wavelength limit $q \to 0$ and on  the regime
of higher frequencies $\omega \sim \omega_p$. Here it follows from the
continuity equation,
\begin{equation}
\chi_0({\bf q}_\parallel,\omega)= {i  q_\parallel^2 \tilde
\sigma(\omega) \over \omega e^2}, \label{eq4}
\end{equation}
where $\tilde\sigma(\omega)$ is the (complex) planar conductivity, which
includes short-range correlation effects.

The optical conductivity $\sigma(\omega)={\rm Re} \tilde\sigma(\omega)
$ in cuprates has been studied extensively \cite{tann},
revealing anomalous behavior consistent with the MFL scenario
\cite{varm,litt}. There have been also numerous theoretical
calculations of $\sigma(\omega)$ at $T=0$, using exact diagonalization
of finite-size models  for strongly correlated electrons in
cuprates \cite{dago}. Here we restrict our attention to
the $t$-$J$ model \cite{rice}, which incorporates the
interplay between antiferromagnetic (AFM) spin correlations governed
by the exchange $J$ and the motion of electrons governed by the
hopping parameter $t$, and the exclusion of double occupation of sites. 
Note that within this
model the only relevant parameters are $J/t$ and the concentration
$c_h$ of holes introduced by doping. 
It is expected that the $t$-$J$ model can represent (even
quantitatively) quite realistic low-energy properties of electrons in
cuprates \cite{jakl3}.  Similar results should also follow for 
the Hubbard model, provided that the strong correlation regime $J\ll
t, U\ll t$ is considered. For a number of observables it is
established that the inclusion of the next-nearest neighbor hopping
$t'$ can improve the agreement with experiments \cite{dago}. However there
are indications that the qualitative behavior of $\sigma(\omega)$ is
less sensitive to the introduction of $t'$ \cite{jakl3}, therefore
we omit such terms in this study.

Recent numerical studies using the finite-temperature Lanczos method
(for a review see \cite{jakl3}) of the $t$-$J$ model with realistic
parameters $J/t=0.3$ reveal results for $\sigma(\omega)$, which are quite
consistent with experiments in the normal state of cuprates. In
particular, it has been found in the intermediate doping regime,
$0.1<c_h<0.3$, that in a broad frequency range $\omega<\omega^*
\sim3t$ $\sigma(\omega)$ follows a novel universal
law \cite{jakl2},
\begin{equation}
\sigma(\omega)=C_0 {1-e^{-\omega/T} \over \omega}, \qquad
\omega<\omega^*. \label{eq5}
\end{equation}
This anomalous diffusion has its origin in the fast decay
of current-current correlations within a disordered spin background
and seems to be  generic for correlated systems \cite{tsun}, at
least at intermediate temperatures $T>T^*$.

On the other hand, in a tight-binding model one can always express the
complex conductivity  $\tilde \sigma(\omega)$ in terms of a memory function
$M(\omega)$ as
\begin{equation}
\tilde \sigma(\omega)= {i e^2{\cal K}t/d \over \omega + M(\omega)},
\qquad \qquad {\cal K} = -{E_{kin} \over 2 N t}, \label{eq6}
\end{equation}
where ${\cal K}$ is the dimensionless kinetic energy density (per
cell) at given $T$.  For the case of a weak scattering $|M(\omega_p)|
\ll \omega_p$, Eqs.(\ref{eq1},\ref{eq2},\ref{eq4},\ref{eq6}) yield
the plasma frequency
\begin{equation}
\omega_p= \sqrt{{e^2 {\cal K}t \over
 \epsilon_0 \epsilon_{\parallel} d}}. \label{eq6a}
\end{equation}
$E_{kin}=\langle H_{kin} \rangle$ is quite well established at $T=0$
from small-system studies \cite{dago}, and it is expected to vary only
slightly at low $T<J$ \cite{jakl3}. Note that at very low doping, e.g.
$c_h<0.05$, assuming independent holes one expects $E_{kin} \propto
c_h$. For higher doping the behavior remains qualitatively
similar, although the slope $d|E_{kin}|/d c_h$ is reduced.  Taking
into account our numerical values for $E_{kin}$ obtained for $J=0.3~t$
and systems with $N=16, 18, 20$ sites as well as the corresponding
number of holes $0<N_h<5$, we get the results for $\omega_p(c_h=N_h/N)$
shown in Fig.~1. In the evaluation of Eq.(\ref{eq6a}) we assume
$t\sim 0.4~$eV \cite{rice} and use parameters relevant for LSCO,
i.e. $\epsilon_{\parallel}=4$ and $d=0.66$~nm \cite{tann,uchi}.

\begin{figure}
\begin{center}
\epsfig{file=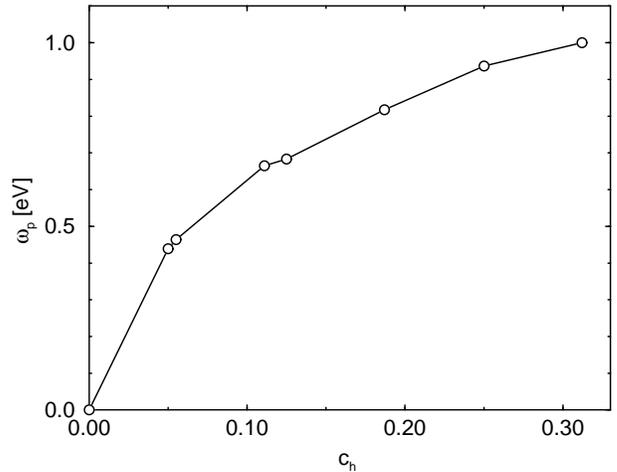,height=8cm,angle=-90}
\end{center}
\caption{ \label{fig1} Plasma frequency $\omega_{p}$ calculated within
the $t$-$J$ model assuming parameters relevant for LSCO: $t=0.4$ eV,
$J=0.3~t$, $\epsilon_{\infty}=4$ and $d =0.66$~nm.  }
\end{figure}

It is evident that Eq.(\ref{eq5}) cannot be modelled
with small $M(\omega)$. It is useful to present $\tilde
\sigma(\omega)$ in the generalized Drude form \cite{tann},
\begin{equation}
\tilde \sigma(\omega) = {i \omega_p^2 \over \eta(\omega)[\omega +i
\gamma(\omega)]}. \label{eq7}
\end{equation}
with an effective relaxation rate (inverse relaxation time)
$\gamma(\omega)$ and the mass renormalization $\eta(\omega)$,
\begin{equation}
{\gamma(\omega)} ={ M''(\omega)\over \eta(\omega) },
\qquad \eta(\omega) = 1+{M'(\omega)\over\omega}. \label{eq8}
\end{equation}
Since $\eta(\omega_p) \neq 1$, it would be more appropriate to define
$\omega^*_p = \omega_p/\sqrt{\eta(\omega^*_p)}$, which should more
directly apply to the $q_z=0$ plasmon resonance, relevant for optical
mesurements and for the transmission EELS. Our numerical $T>0$ results
for $\sigma(\omega)$ and consequently for $M(\omega)$
\cite{jakl2,jakl3} reveal however quite consistently for doping concentrations
$0.05<c_h<0.3$ that $\eta(\omega_p)\alt 1$, hence $\omega^*_p \sim
\omega_p$.

A peculiar feature of the charge dynamics in cuprates is the non-Drude
behavior of $\sigma(\omega)$, as e.g. represented by the MFL scenario
\cite{varm}, or more specifically by Eq.(\ref{eq5}), which should be
valid at least near optimal doping.  Within the
broader MFL concept the dependence $\gamma(\omega,T)$
\cite{litt,jakl1} is given by,
\begin{equation}
\gamma = \tilde \lambda (\omega + \xi T), \qquad \tilde \lambda = 2\pi
\lambda. \label{eq9}
\end{equation}
This behavior is obeyed in cuprates up to $\omega \agt \omega_p$.
The dimensionless constant $\tilde\lambda$ obtained from numerical studies
is not small. Estimates for various compounds fall in the range $0.5<\tilde
\lambda<0.9$ \cite{jakl3}.

Although $\tilde \lambda$ appears as a free parameter within the MFL
proposal, this is not the case with Eq.(\ref{eq5}), which also
exhibits the MFL variation, Eq.(\ref{eq9}), apart from logarithimic
corrections. Assuming that $\sigma(\omega>\omega^*) =0$, one can
analytically evaluate $\tilde \sigma(\omega)$ and consequently
$M(\omega)$. It is easy to express $\gamma$ in two regimes:

\noindent a) For $\omega \ll T \ll \omega^*$ the expansion in $\omega$
yields
\begin{equation}
\tilde \lambda = {\pi \over 3 + 2{\rm ln}T/\omega}, \qquad \xi =2,
\label{eq10}
\end{equation}
\noindent b) while for $T \ll \omega \ll \omega^*$ we get
\begin{equation}
\tilde \lambda = {\pi \over 2(1 + {\rm ln}\omega/T)}.
\label{eq10a}
\end{equation}
For the relevant range of $T$ in the normal state of cuprates and the
experimental range of $\omega$ (with reliable results),
Eq.(\ref{eq10a}) should apply to experiments. The best overall fit
of the numerical results
\cite{jakl1,jakl3} for $T\ll t$ and $\omega<t$ is achieved
with $\tilde \lambda \sim 0.6$ and $\xi \sim 2.7$, also quite
consistent with experiments \cite{tann}.

If we insert the generalized Drude expression (\ref{eq7}) and
Eq.(\ref{eq2}), into Eqs.(\ref{eq4},\ref{eq1}), we get
\begin{equation}
\epsilon({\bf q},\omega)= 1- {\tilde \omega_p^2({\bf q}) \over
\eta(\omega^2 +i\omega \gamma)} ,\label{eq11}
\end{equation}
where $\tilde \omega_p({\bf q})$ is the effective plasmon frequency
for a layered system,
\begin{equation}
\tilde \omega_p ({\bf q}) = \omega_p {q_{\parallel} \over
\sqrt{q^2_{\parallel} + (\epsilon_{\perp}
q_z/\epsilon_{\parallel})^2} }. \label{eq12}
\end{equation}
Note that for $q_{\parallel} < q_z$ the plasmon shows an acoustic
dispersion, i.e. $\tilde \omega_p = \omega_p
q_{\parallel}\epsilon_{\parallel}/q_z \epsilon_{\perp} \propto
q_{\parallel}$ \cite{fett,grif}.

With Eq.(\ref{eq11}) the  EEL function (\ref{eq3})
can be written as
\begin{equation}
I(\omega)= {\gamma \eta \omega \tilde \omega_p^2 \over (\eta \omega^2
- \tilde \omega_p^2)^2 +\gamma^2 \eta^2\omega^2}. \label{eq13}
\end{equation}
Let us restrict our attention again to the intermediate doping regime
where relations (\ref{eq5},\ref{eq9}) apply. In the range $\omega > T$,
mostly relevant to experiments and for the study of the plasma
resonance in particular, we insert $\gamma \sim \tilde \lambda \omega$
and assume $\eta(\omega) \sim 1$.  Note that these simplifications
could be possibly violated in the extreme acoustic limit $\tilde
\omega_p({\bf q}) \to 0$.  The result is
\begin{equation}
I(\omega)= {\tilde \lambda x^2 \over (x^2 -1)^2 + \tilde \lambda^2
x^4},\qquad x={\omega \over \tilde\omega_p}.
\label{eq14}
\end{equation}
It is evident from  expression  (\ref{eq14})  that  the width  of  the
plasmon resonance is entirely determined by $\tilde \lambda$, i.e. the
resonance  width    is $\Delta   \omega/\tilde \omega_p   \sim  \tilde
\lambda$.  A further characteristic feature   is the anomalous  variation
below  the   resonance,   i.e.   $I(\omega<\tilde  \omega_p)   \propto
\omega^2$. 

\begin{figure}
\begin{center}
\epsfig{file=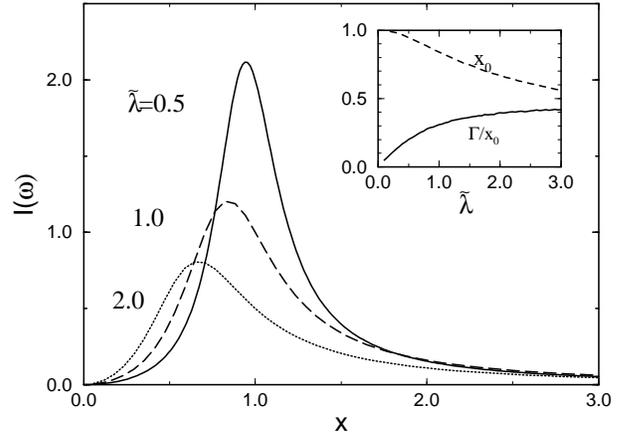,width=8cm}
\end{center}
\caption{ \label{fig2} Loss function $I$ (Eq.[16])
vs. $x=\omega/\tilde \omega_p$ for three different
$\tilde\lambda$-values, i.e. $\tilde\lambda$=0.5, 1.0 and 2.0.  The
inset shows the ratio $\Gamma/x_0$ (solid curve) between the
half-width $\Gamma$ at  half-maximum measured at the low-energy side of
the peak and the renormalized plasma frequency
$x_0=\omega_{max}/\tilde \omega_{p}$ (dashed curve).  }
\end{figure}

Figure~2 presents $I(\omega)$  for three different values of
$\tilde   \lambda$.  For a typical  value   $\tilde\lambda\sim 1$ this
implies that the plasmon resonance is necessarily very broad, i.e. the
halfwidth  of the  peak of $I(\omega)$  is of the  order of the
plasma  frequency itself.   This  seems to  be  a new  aspect  of  the
anomalous behavior of strongly correlated  electrons in doped AFM and
cuprates.   In addition, there  is  a noticeable downward
shift of the peak position with respect to $\tilde\omega_p$.

Let us turn to the discussion of experimental results of cuprates.
The loss function $I({\bf q},\omega)$ has been determined directly by
EELS measurements for BSCCO \cite{nuck,wang} and for
YBa$_2$Cu$_3$O$_7$\cite{Tarrio88}.  The characteristic asymmetric
shape of $I(\omega)$ and the strong damping of the plasmon peak,
emerging from our analysis, appears to be consistent with these
experiments

The determination of the loss function $I(\omega)$ from optical $q \to
0$ data turns out to be also difficult, in particular because of
the limited spectral range of the reflectivity data which makes the
Kramers-Kronig transformation less reliable.  However the use of
ellipsometric data appears to provide highly accurate results, which
are fully consistent with the existing EELS data
\cite{Bozovic90,Kim91}. $I(\omega)$ determined for a number of
high-T$_c$ superconductors show a single broad peak in the energy
range below 2 eV.  A characteristic measure for the width of the peak,
which can be easily compared with experimental data, is the half-width
$\Gamma$ taken at the low-energy side of the peak at $x_0$.
For $\tilde \lambda=1.0$ we obtain $\Gamma/x_0\sim 0.31$
(see inset of Fig. 2).  Typical values extracted from experimental
loss function data are 0.35 for BSCCO, 0.40 for
YBa$_2$Cu$_3$O$_7$, and 0.45 for Tl$_2$Ba$_2$Ca$_2$Cu$_3$O$_{10}$,
i.e. consistent with the theoretical values for $\Gamma/x_0$.
Furthermore it has been noted that the low-energy part of the loss
function of all systems is quadratic in frequency,
i.e. $I(\omega)\propto \omega^2$, nearly up to the plasma
frequency. It has been conjectured that this quadratic frequency
dependence is universally obeyed in all high-T$_c$ superconductors
\cite{Bozovic90,Kim91}, although at lowest frequencies ($\omega < 0.2$
eV) the dependence seems to be linear\cite{Humlicek92}.  We note that
$I(\omega<\omega_p)$, as given by Eq.(\ref{eq13}), shows indeed an
approximative quadratic dependence, which is due to the anomalous
damping, Eq.(\ref{eq9}), while at low frequency the
$\omega$-dependence becomes linear, because of the $T$-scale in
Eq. (\ref{eq9}).

As we have assumed equidistant planes, we shall focus here on the
single-layer material LSCO for a quantitative comparison between
theory and experiment. For moderate doping the plasmon peak in the
loss function is found at 0.80 (0.83) eV for $x=0.10~ (0.20)$,
respectively \cite{uchi}.  These values and $\Gamma/x_0\sim 0.3$ are in
reasonable agreement with our numerical results.  It is surprising,
however, that the experimental plasma frequency is not monotonous in
its doping dependence. For $x=0.34$ Uchida {\it et al.}\cite{uchi}
report $\omega_p=0.77$~eV. This discrepancy may be related to the interplay with
interband excitations in the highly doped material, which are not
contained in the $t$-$J$ model. Yet it is remarkable that in a model
where the physics scales with the hole-concentration (which is rather
small) also the correct scale of the plasma oscillation can be
obtained\cite{density}.

In summary we have shown, that the anomalous damping of the optical
conductivity, as obtained within the $t$-$J$ model, explains the
universal line shape of the loss function observed in the optimal
doping range\cite{Bozovic90}.  In this regime the width of the plasmon
peak is of the order of the plasmon energy itself. Model results give
besides the qualitative also a good quantitative description of EEL and the
plasmon resonance in cuprates, at least in the intermediate doping
regime.


\begin{references}

\bibitem{tann} D. B. Tanner and T. Timusk, in {\it Physical Properties
of High Temperature Superconductors III}, ed. by D. M. Ginsberg (World
Scientific, Singapore, 1992), p.363.
\bibitem{pine} D. Pines and P. Nozieres, {\it The theory of quantum
liquids}, Vol. I, Benjamin, (1966).
\bibitem{nuck} N. N\"ucker, H. Romberg, S. Nakai, B. Scheererm
J. Fink, Y. F. Yan, and Z. X. Zhao, Phys. Rev. B {\bf 39}, 12379
(1989).
\bibitem{wang} Y.-Y.Wang, G. Feng, and A. L. Ritter, Phys. Rev. B {\bf
42}, 420 (1990).
\bibitem{uchi} S. Uchida, T. Ido, H. Takagi, T. Arima, Y. Tokura, and
S. Tajima, Phys. Rev. B {\bf 43}, 7942 (1991).
\bibitem{schu} K. Schulte, private communication.
\bibitem{fett} A. L. Fetter, Ann. Phys. {\bf 88}, 1 (1974).
\bibitem{grif} A. Griffin, Phys. Rev. B {\bf 37}, 5943 (1988);
A. Griffin and A. J. Pindor, Phys. Rev. B {\bf 39}, 5943 (1989).
\bibitem{legg} A. J. Leggett, J. Phys. Chem. Solids {\bf 59},
1729 (1998).
\bibitem{varm} C. M. Varma, P. B. Littlewood, S. Schmitt-Rink,
E. Abrahams, and A. E. Ruckenstein, Phys. Rev. Lett. {\bf 63}, 1996
(1989).
\bibitem{litt} P. B. Littlewood, in {\it Proceedings of the Les
Houches Summer School, Session LVI},  Eds. B. Doucot and
J. Zinn-Justin (Elsevier, Amsterdam), p.69 (1995).
\bibitem{jakl1} J. Jakli\v c and P. Prelov\v sek, Phys. Rev. B {\bf
52}, 6903 (1995).
\bibitem{jakl2} J. Jakli\v c and P. Prelov\v sek, Phys. Rev. Lett. {\bf
75}, 1340 (1995).
\bibitem{jakl3} J. Jakli\v c and P. Prelov\v sek, to appear in
Adv. Phys.
\bibitem{khal} G. Khaliullin and P. Horsch, Phys. Rev. B {\bf 54},
R9600 (1996).
\bibitem{tohy} T. Tohyama, P. Horsch, and S. Maekawa,
Phys. Rev. Lett. {\bf 74}, 980 (1995); R. Eder, Y. Ohta, and
S. Maekawa, Phys. Rev. Lett. {\bf 74}, 5124 (1995).
\bibitem{rice} T. M. Rice, in {\it Proceedings of the Les
Houches Summer School, Session LVI}, Eds. B. Doucot and
J. Zinn-Justin (Elsevier, Amsterdam), p.19 (1995).
\bibitem{dago} E. Dagotto, {\it Rev. Mod. Phys.} {\bf 66}, 763 (1994).
\bibitem{tsun} H. Tsunetsugu and M. Imada, J. Phys. Soc. Jpn. {\bf
66}, 1876 (1997).
\bibitem{Tarrio88} C. Tarrio and S. E. Schnatterly, Phys. Rev. B {\bf
38}, 921 (1988).
\bibitem{Bozovic90} I. Bozovic, Phys. Rev. B {\bf 42}, 1969 (1990).
\bibitem{Kim91} J. H. Kim {\it et al.}, Physica C {\bf 185-189}, 1019 (1991).
\bibitem{Humlicek92} J. Humli\v cek, J. Kircher, H.-U. Habermeier, and
M. Cardona, Physica C {\bf190}, 383 (1992).
\bibitem{density}
Note that by using the standard form for the plasma frequency
$\omega_p^{\infty}=\sqrt{\epsilon_{\infty}}\; \omega_p=\sqrt{n
e^2/\epsilon_0 m^*}$ much higher carrier concentrations are usually
inferred\cite{Bozovic90}.
\end{references}
\end{document}